\documentclass{tcibook}
\usepackage{fancyhea}
\usepackage{work}
\usepackage{bm}       
\usepackage{graphicx}
\usepackage{hyperref}      
\usepackage{color}
\usepackage{amssymb,amsmath}

\newcommand{\nc}{\newcommand}  

\def\beq{\begin{equation}}
\def\eeq#1{\label{#1}\end{equation}}
\def\eeqn{\end{equation}}

\newenvironment{Eqnarray}%
   {\arraycolsep 0.14em\begin{eqnarray}}{\end{eqnarray}}
\def\beqa{\begin{Eqnarray}}
\def\eeqa#1{\label{#1}\end{Eqnarray}}
\def\eeqan{\end{Eqnarray}}

\nc{\ra}{\rightarrow}  
\nc{\slsh}{\slash\hspace*{-0.22cm}}
\def\Re{{\cal R \mskip-4mu \lower.1ex \hbox{\it e}\,}}
\def\Im{{\cal I \mskip-5mu \lower.1ex \hbox{\it m}\,}}

\nc{\vev}[1]{ \left\langle {#1} \right\rangle }
\nc{\bra}[1]{ \langle {#1} | }
\nc{\ket}[1]{ | {#1} \rangle }
\nc{\fb}{\,{\rm fb}^{-1}}
\nc{\ev}{{\rm eV}}
\nc{\kev}{{\rm keV}}
\nc{\Mev}{{\rm MeV}}
\nc{\gev}{{\rm GeV}}
\nc{\tev}{{\rm TeV}}
\nc{\mev}{{\rm MeV}}

\def\del{\partial}
\def\Dslash{\not{\hbox{\kern-4pt $D$}}}
\def\dslash{\not{\hbox{\kern-2pt $\del$}}}
\def\pslash{\not{\hbox{\kern-2pt $p$}}}
\def\ETmiss{ \not{\hbox{\kern-4pt $E$}}_T }

\def\BR{\mbox{\rm BR}}

\def\msb{{\bar{\ssstyle M \kern -1pt S}}}

\begin{document}

\def\bibname{References}
\bibliographystyle{plain}

\raggedbottom

\pagenumbering{roman}

\parindent=0pt
\parskip=8pt
\setlength{\evensidemargin}{0pt}
\setlength{\oddsidemargin}{0pt}
\setlength{\marginparsep}{0.0in}
\setlength{\marginparwidth}{0.0in}
\marginparpush=0pt


\pagenumbering{arabic}

\renewcommand{\chapname}{chap:intro_}
\renewcommand{\chapterdir}{.}
\renewcommand{\arraystretch}{1.25}
\addtolength{\arraycolsep}{-3pt}

 
\chapter{Lattice field theory for the energy and intensity frontiers:  \\Scientific goals and computing needs}
\label{chap:LFT}

\begin{center}\begin{boldmath}

\begin{center}

\begin{large} {\bf Conveners: T. Blum (U. Connecticut), R. S. Van de Water (FNAL)\\ Observer: D. Holmgren (FNAL)} \end{large}

R.~Brower~(Boston~U.),
S.~Catterall~(Syracuse~U.),
N.~Christ~(Columbia~U.),
A.~Kronfeld~(FNAL),
J.~Kuti~(UCSD),
P.~Mackenzie~(FNAL),
E.~T.~Neil~(U.~Colorado),
S.~R.~Sharpe~(U.~Washington),
R.~Sugar~(UCSB)

\end{center}

\end{boldmath}\end{center}


\section{Introduction}
\label{sec:lqcd:intro}

One of the foremost goals of high-energy physics is to test the
Standard Model of particle physics and to search for indications of
new physics beyond.  Towards this aim, the experimental high-energy
physics program is pursuing 
three complementary approaches: experiments
at the ``energy frontier" try to directly produce non-Standard Model
particles in collisions at large center-of-mass energies;
experiments at the ``cosmic frontier" look for
  astronomical evidence of new interactions and aim to detect
  cosmically-produced non-Standard-Model particles through their interaction with
  ordinary matter; while experiments at the ``intensity
frontier"~\cite{Hewett:2012ns} make precise measurements of rare
processes and look for discrepancies with
the Standard Model.   In many cases, interpretation of the experimental measurements requires a quantitative of understanding the nonperturbative dynamics of the quarks and gluons in the underlying process.  Lattice gauge theory provides the only known method
for \emph{ab initio} quantum chromodynamics (QCD) calculations with controlled uncertainties, by casting
the fundamental equations of QCD into a form amenable to high-performance
computing.  Thus, facilities for numerical lattice QCD are an
essential theoretical adjunct to the experimental high-energy physics
program.  This report describes the computational and software infrastructure resources needed for lattice gauge theory to meet the scientific goals of the future energy- and intensity-frontier experimental programs.  We focus on the efforts and plans in the US, but comparable efforts are ongoing in Europe and Japan.

Experiments at the intensity frontier probe
quantum-mechanical loop effects; thus they can be sensitive to physics at
higher energy scales than those directly accessible at the LHC.  Measurements in the quark-flavor sector, for example, constrain the scale of new particles with ${\mathcal O}(1)$ couplings to be greater than 1,000~TeV or even 10,000~TeV~\cite{Isidori:2010kg}.  
Contributions from new heavy particles may be
observable as deviations of the measurements from Standard-Model
expectations, provided both the experimental measurements and
theoretical predictions are sufficiently precise.  For many quantities, the comparison between the measurements and Standard-Model predictions are currently limited by theoretical uncertainties from nonperturbative hadronic amplitudes such as decay constants, form 
factors, and meson-mixing matrix elements.  These nonperturbative hadronic parameters can only be calculated with controlled uncertainties that are systematically improvable using numerical lattice QCD.  The U.S. Lattice-QCD Collaboration (USQCD) lays out an ambitious five-year vision for future lattice-QCD calculations in the white paper ``Lattice QCD at the Intensity Frontier"~\cite{USQCD_IF_whitepaper13}, explaining how they can provide essential and timely information for current, upcoming, and proposed experiments.  
In some cases, such as for the determination of CKM matrix elements that are parametric inputs to
Standard-Model predictions, improving the precision of existing calculations is sufficient, and the expected
increase in computing power due to Moore's law will enable a continued reduction in errors.
In other cases, like the muon $g-2$ and the nucleonic probes of non-Standard-Model physics, new hadronic matrix elements
are required; these calculations are typically computationally more demanding, and methods are
under active development.  

Precision measurements at high-energy colliders can also probe quantum-mechanical loop effects.  Future proposed facilities such as the International Linear Collider (ILC), Triple-Large Electron-Positron Collider (TLEP), or a muon collider would enable dramatic improvements in measurements of Higgs partial widths to sub-percent precision, but a reduction in the theoretical uncertainties on the Standard-Model predictions to this level will be needed to fully exploit these measurements.  Currently parametric errors from the quark masses $m_c$ and $m_b$ and the strong coupling constant $\alpha_s$ are the largest sources of uncertainty in the Standard-Model branching-ratio predictions for the dominant Higgs decay mode $H \to \bar{b}b$, many other Higgs decay channels, and the Higgs total width~\cite{Denner:2011mq}.  Numerical lattice-QCD simulations provide the only first-principles method for calculating the parameters of the QCD Lagrangian; in fact, they currently provide the single-most precise determinations of $\alpha_s$ and $m_c$, and a competitive calculation of $m_b$.  In the next few years, increased computing resources will enable a significant reduction in the uncertainty on $m_b$, smaller reductions in the errors on $m_c$ and $\alpha_s$, and further corroboration for all of these quantities from independent lattice calculations.  Ultimately, the goal is to improve the accuracy of QCD calculations to the point where they no longer limit what can be learned from high-precision experiments at both the energy and intensity frontiers that seek to test the Standard Model.  Indeed, lattice-QCD calculations of hadronic matrix elements and fundamental QCD parameters may play a key role in definitively establishing the presence of physics beyond-the-Standard Model and in determining its underlying structure.

The Large Hadron Collider (LHC) aims to directly produce new particles in high-energy proton-proton collisions that can be detected by the ATLAS and CMS experiments, and already these experiments have discovered a $\sim$ 125 GeV Higgs-like particle~\cite{Aad:2012tfa,Chatrchyan:2012ufa}.   If, however, electroweak symmetry breaking is realized in Nature via the Standard-Model Higgs, the mass of the light Higgs must be finely tuned, leading to the well-known hierarchy problem.   Therefore many proposed new-physics models aim to provide a deeper dynamical mechanism that resolves this shortcoming of the Standard Model.  Examples include technicolor~\cite{Farhi:1980xs,Hill:2002ap}, in which the Higgs may be a dilaton associated with the breaking of conformal ({\it i.e.} scale) symmetry, or little Higgs scenarios~\cite{Kaplan:1983sm,ArkaniHamed:2002pa,ArkaniHamed:2002qy}, in which the Higgs is a pseudo-Goldstone boson associated with chiral-symmetry breaking.  The common thread in these classes of new-physics models is that the Higgs boson is composite, and its dynamics are nonperturbative near the electroweak scale.  Therefore a natural tool for studying these theories is lattice gauge theory.   In recent years, members of the lattice-gauge-theory community have been developing methods to study gauge theories beyond QCD.  The USQCD Collaboration documents progress in lattice calculations for beyond-the-Standard Model physics in the white paper ``Lattice Gauge Theories at the Energy Frontier"~\cite{USQCD_EF_whitepaper13}, and outlines strategic goals for the next five years focusing on aiding new-physics searches at the LHC.  The current highest priority is to find a viable composite-Higgs model with a light scalar and an oblique $S$-parameter~\cite{Peskin:1990zt} consistent with precision electroweak constraints, and then to compute predictions of this theory that can be tested at the 14-TeV LHC run for other quantities such as the heavier particle spectrum and $W$-$W$ scattering.  More broadly, the goal of the lattice beyond-the-Standard Model effort is to develop quantitative tools for studying new gauge theories, which may behave quite differently than naive expectations based on intuition from QCD.  In the future, numerical lattice gauge theory calculations can provide essential quantitative input to Higgs (and other new-physics) model building, and, ultimately, play a key role in discovering TeV-scale strong dynamics. 

The lattice gauge theory research community in the United States coordinates much of its effort to obtain
computational hardware and develop software infrastructure through the USQCD Collaboration.
Support for USQCD has been obtained from the high-energy physics and nuclear physics offices of DOE in the
form of (i) funds for hardware and support staff, (ii) computational resources on leadership-class machines
through INCITE awards, and (iii) SciDAC awards for software and algorithm development.
The first has consisted of two 4--5 year grants, the second of which extends until 2014.
Since its inception, the INCITE program has awarded computing resources to USQCD every year.
SciDAC has funded three software projects for lattice QCD, the most recent beginning in 2012.
All three components have been critical for progress in lattice QCD in the past decade.
The primary purpose of USQCD is to support the high-energy and nuclear physics experimental programs in the
U.S. and worldwide.
To this end, USQCD establishes scientific priorities, which are documented in white papers~\cite{USQCD_EF_whitepaper13,USQCD_IF_whitepaper13,USQCD_NP_whitepaper13,USQCD_Thermo_whitepaper13}.
USQCD's internal and INCITE computing resources are then allocated, in a proposal driven process, to
self-directed smaller groups within USQCD to accomplish these goals.

At present, members of USQCD make use of dedicated high-capacity PC and GPU cluster funded by the DOE through the LQCD-ext
Infrastructure Project, as well as a Cray XE/XK computer, and IBM Blue Gene/Q and Blue Gene/P computers, made
available by the DOE's INCITE Program.
During 2013, USQCD as a whole expects aggregate production of 214 sustained Tflop/sec-yrs from these machines, where "sustained Tflop/sec" refers to the floating point performance of lattice-QCD codes.
USQCD also has a PRAC grant to develop code for the NSF's new petascale computing facility, Blue Waters.
Further, subgroups within USQCD apply individually to utilize other DOE and NSF supercomputer centers.
For some time, the resources USQCD has obtained have grown with a doubling time of approximately 1.5~years,
consistent with Moore's law, and this growth rate will need to continue to meet the collaboration's scientific objectives.
 
The software developed by USQCD under a SciDAC grant enables U.S. lattice gauge theorists to use a wide variety
of architectures with very high efficiency, and it is critical that USQCD software efforts continue at their
current pace.
Historically, the advance preparation of USQCD for new hardware has enabled members to take full advantage of
early science time that is often available while new machines are coming online and being tested.
Over time, the development of new algorithms has had at least as important an impact on the field of lattice
QCD as advances in hardware, and this trend is expected to continue, although the rate of algorithmic
advances is not as smooth or easy to predict as that of hardware.

This report presents the future computing needs and plans of the U.S. lattice gauge theory community, and argues that continued support of the U.S. (and worldwide) lattice-QCD effort is essential to fully capitalize on the enormous investment in the high-energy physics experimental program.  This report is organized as follows.  Section~\ref{sec:lqcd:physics} presents the role of lattice QCD to aid in the search for new physics at the energy and intensity frontiers.  Next, Section~\ref{sec:lqcd:resources} presents details of the computational hardware and software resources needed to undertake
these calculations.   Finally, Section~\ref{sec:lqcd:summ} recaps the main findings of this report.   Achieving the scientific goals outlined in the USQCD white papers~\cite{USQCD_EF_whitepaper13,USQCD_IF_whitepaper13} and summarized here will require support of both the national supercomputing centers and of dedicated USQCD hardware, investment in software development, and support of postdoctoral researchers and junior faculty.  Given sustained investment in numerical lattice field theory, the lattice community will continue to carry out the nonperturbative theoretical calculations needed to support the current and future experimental particle-physics programs at the energy and intensity frontiers.

\section{Physics motivation}
\label{sec:lqcd:physics}

In this section we first provide a brief introduction to lattice QCD.  We summarize the dramatic progress in the past decade, with some emphasis on calculations carried out under the auspices of USQCD, and highlight calculations that validate the whole paradigm of numerical lattice-QCD.  This sets the stage for Secs.~\ref{subsec:lqcd:IF} and~\ref{subsec:lqcd:EF}, which describe a broad program of lattice-QCD calculations that will be relevant for experiments at the intensity and energy frontiers, respectively.  

\subsection{Lattice field theory methodology and validation}
\label{subsec:lqcd:validation}

Lattice gauge theory formulates QCD on a discrete Euclidean spacetime lattice, thereby transforming the
infinite-dimensional quantum field theory path integral into a finite-dimensional integral that can be solved
numerically with Monte Carlo methods and importance sampling.
In practice, lattice-QCD simulations are computationally intensive and require the use of the world's most
powerful computers.
The QCD Lagrangian has $1 + N_f + 1$ parameters: the gauge coupling $g^2$, the $N_f$ quark masses $m_f$, and 
the $CP$-violating parameter $\bar{\theta}$.
Because measurements of the neutron electric dipole moment (EDM) bound $\bar{\theta} < 10^{-10}$, most
lattice-QCD simulations set $\bar{\theta} = 0$.
The gauge-coupling and quark masses in lattice-QCD simulations are tuned by calibrating to $1 + N_f$
experimentally-measured quantities, typically hadron masses or mass-splittings.
Once the parameters of the QCD action are fixed, everything else is a prediction of QCD.

There are many ways to discretize QCD, particularly the fermions, but all of them recover QCD in the
continuum limit, i.e., when the lattice spacing $a\to 0$.
The various fermion formulations in use have different advantages (such as computational speed or exact
chiral symmetry) and different sources of systematic uncertainty; hence it is important to compute quantities
with more than one method for independent validation of results.
The time required for numerical simulations increases as the quark mass decreases (the condition number of
the Dirac operator, which must be inverted, increases with decreasing mass), so quark masses in lattice
simulations have usually been higher than those in the real world.
Typical lattice calculations now use quark masses such that the pion mass $m_\pi \lesssim 300$~MeV, while 
state-of-the art calculations for some quantities attain pions at or slightly below the physical mass of 
$m_\pi\sim140$~MeV. 
Over the coming decade, improvements in algorithms and increases in computing power will render chiral
extrapolations unnecessary.

Most lattice-QCD simulations proceed in two steps.
First one generates an ensemble of gauge fields with a distribution $\exp[-S_\text{QCD}]$; next one
computes operator expectation values on these gauge fields.
A major breakthrough in lattice-QCD occurred with the advent of gauge-field ensembles that include the
effects of the dynamical $u$, $d$, and $s$ quarks in the vacuum.
Lattice-QCD simulations now regularly employ ``$N_f = 2+1$" sea quarks in which the light $u$ and $d$
sea-quark masses are degenerate and somewhat heavier than the physical values, and the strange-sea quark mass
takes its physical value.
Further, ``$N_f = 2 + 1 + 1$" simulations that include a charm sea quark are now underway; dynamical charm
effects are expected to become important as precision for some quantities reaches the percent level.
During the coming decade, even $N_f=1+1+1+1$ simulations which include isospin-breaking in the sea are
planned.

The easiest quantities to compute with controlled systematic errors and high precision in lattice-QCD
simulations have only a hadron in the initial state and at most one hadron in the final state, where the
hadrons are stable under QCD (or narrow and far from threshold).
These quantities, often referred to as ``gold-plated,'' include meson masses and decay constants,
semileptonic and rare decay form factors, and neutral meson mixing parameters, and enable determinations of
all CKM matrix elements except $|V_{tb}|$.
Many interesting QCD observables are not gold-plated, however, such as resonances like the $\rho$ and $K^*$
mesons, fully hadronic decay matrix elements such as for $K \to \pi\pi$ and $B\to DK$, and long-distance
dominated quantities such as $D^0$-$\bar{D}^0$~mixing.
That said, lattice QCD with current resources is beginning to tackle such quantities, particularly in 
$K\to\pi\pi$ decay.

\begin{figure}
    \centering
    \includegraphics[width=\linewidth]{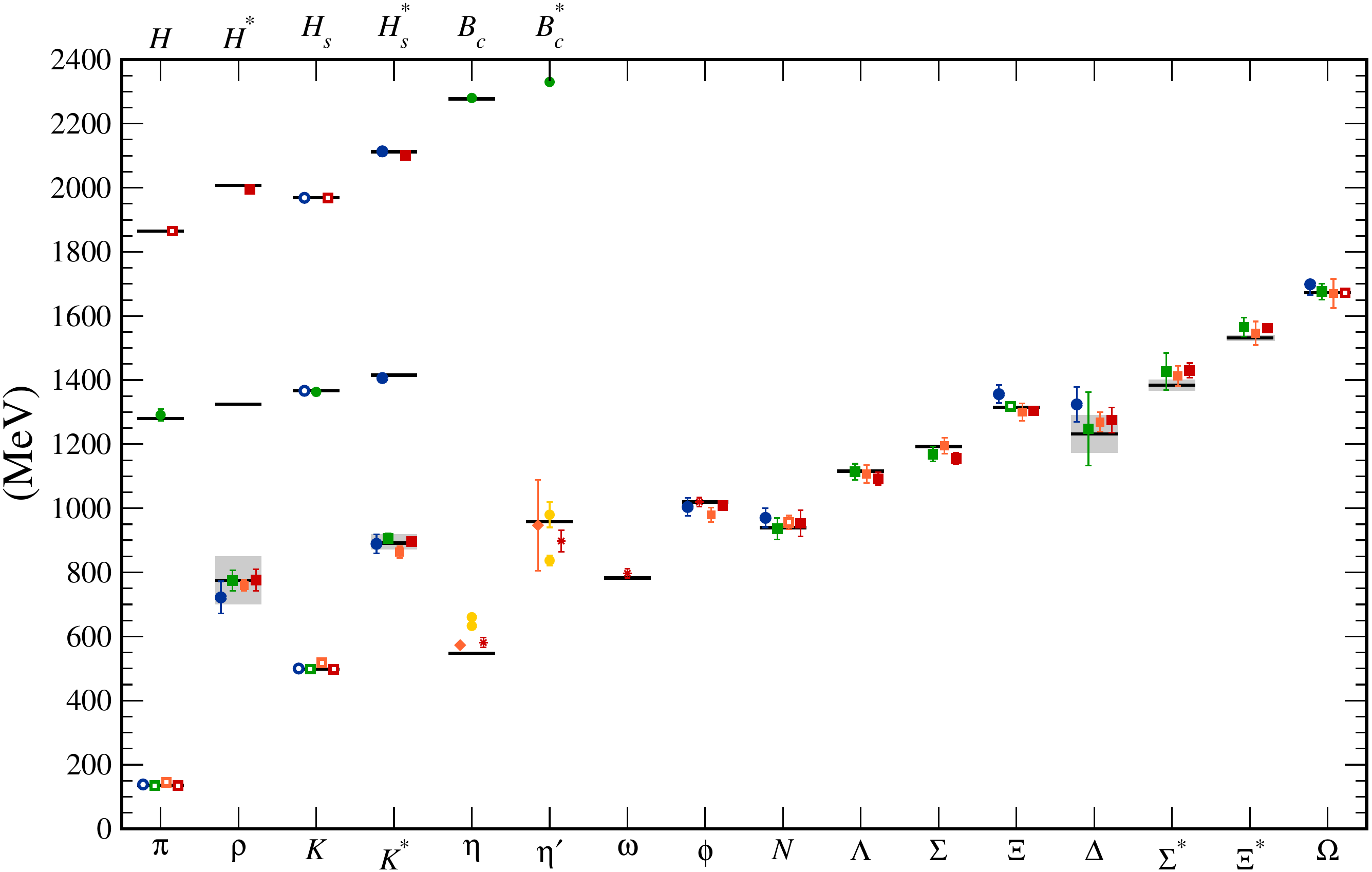}
    \caption[Hadron spectrum from many different lattice-QCD calculations]{Hadron spectrum from many 
        different lattice-QCD calculations~\cite{Aubin:2004wf,Aoki:2008sm,Durr:2008zz,Bazavov:2009bb,Christ:2010dd,Bernard:2010fr,Gregory:2010gm,Dudek:2011tt,Bietenholz:2011qq,Mohler:2011ke,Gregory:2011sg}.
        Open symbols denote masses used to fix bare parameters; closed symbols represent \emph{ab initio}
        calculations.
        Horizontal black bars (gray boxes) show the experimentally measured masses (widths).
        $b$-flavored meson masses ($B_c^{(*)}$ and $H_{(s)}^{(*)}$ near 1300~MeV) are offset by $-4000$~MeV.
        Circles, squares and diamonds denote staggered, Wilson and domain-wall fermions, respectively.
        Asterisks represent anisotropic lattices ($a_t/a_s<1$).
        Red, orange, yellow and green and blue signify increasing ensemble sizes (i.e., increasing range of 
        lattice spacings and quark masses).
        From Ref.~\cite{Kronfeld:2012uk}.}
    \label{lqcd:fig:spectrum}
\end{figure}

\begin{figure}
    \centering
    \includegraphics[width=0.495\textwidth]{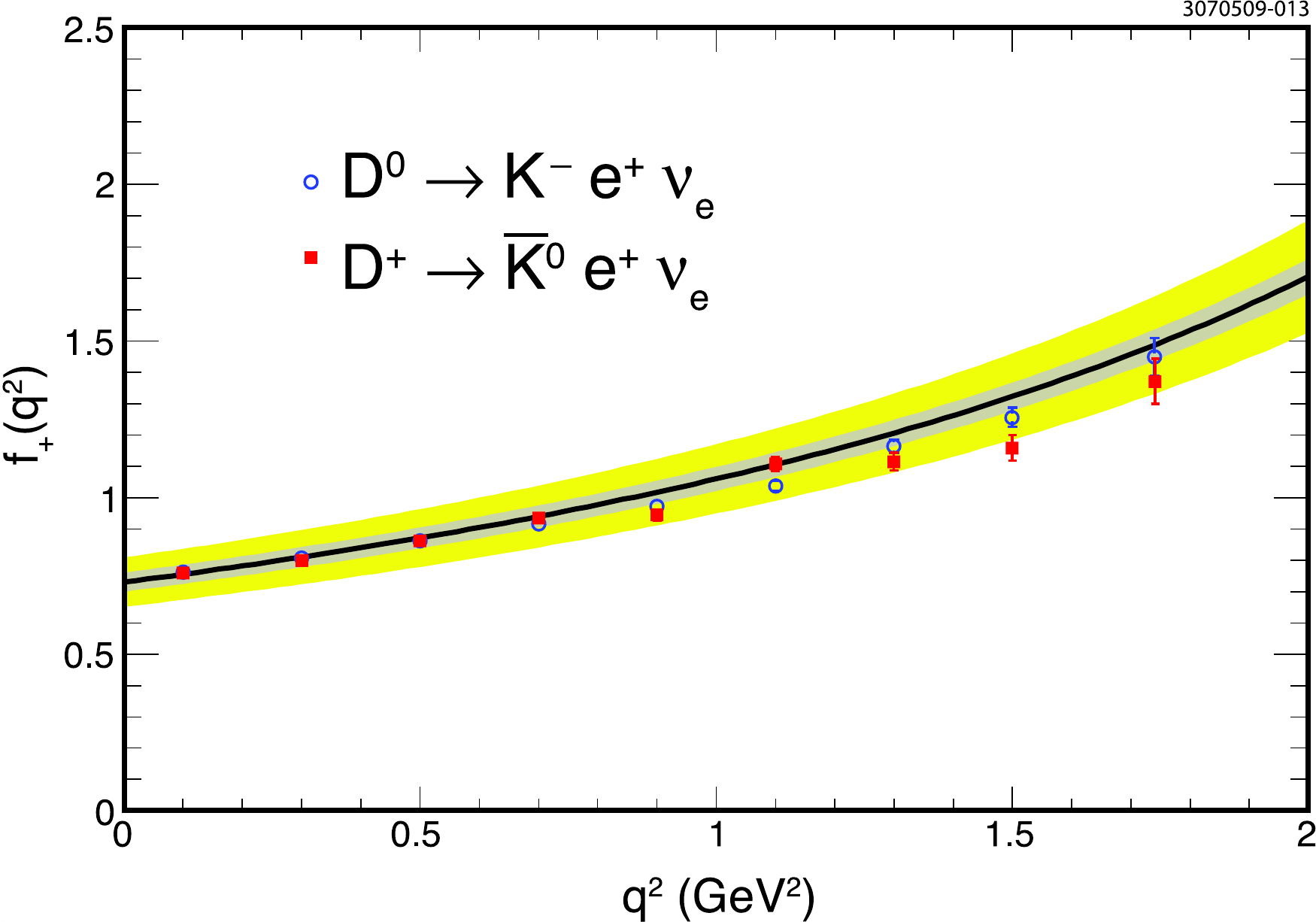}\hfill
    \includegraphics[width=0.495\textwidth]{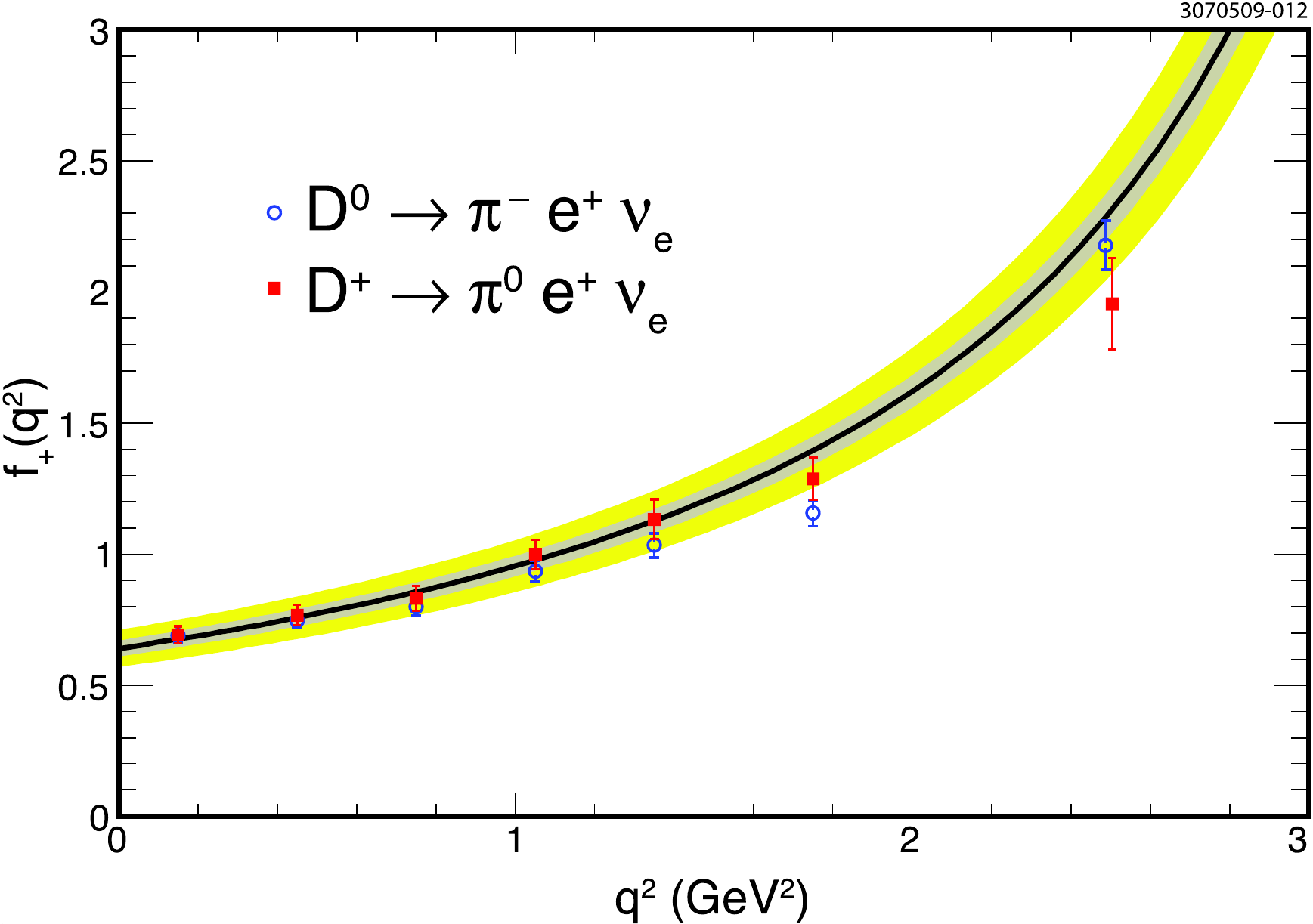}
    \caption[Lattice-QCD calculations of $D$-meson form factors compared with measurements]{Comparison of 
        $N_f = 2+1$ lattice-QCD calculations of $D$-meson form factors~\cite{Aubin:2004ej,Bernard:2009ke} 
        (curves with error bands) with measurements from CLEO~\cite{Besson:2009uv} (points with error bars).
        From Ref.~\cite{Besson:2009uv}.}
    \label{lqcd:fig:D2K}
\end{figure}

Many errors in lattice-QCD calculations can be assessed within the framework of effective field theory.
Lattice-QCD calculations typically quote the following sources of uncertainty:
\begin{itemize}
\item \emph{Monte Carlo statistics and fitting};
\item \emph{tuning lattice spacing and quark masses} by calibrating to a few experimentally-measured 
quantities such as $m_\pi$, $m_K$, $m_{D_s}$, $m_{B_s}$, $m_\Omega$, and $f_\pi$;
\item \emph{matching lattice gauge theory to continuum QCD} using fixed-order lattice perturbation theory, 
step-scaling, or other partly or fully nonperturbative methods;
\item \emph{chiral and continuum extrapolation} by simulating at a sequence of light (up and down) quark
masses and lattice spacings and extrapolating to $m_{\rm lat} \to m_{\rm phys}$ and $a\to0$ using functional
forms derived in chiral and weak-coupling QCD perturbation theory;
\item \emph{finite volume corrections}, which may be estimated using effective theory and/or studied directly 
by simulating lattices with different spatial volumes.
\end{itemize}
The methods for estimating uncertainties can be verified by comparing results for known quantities with
experiment.
Lattice-QCD calculations successfully reproduce the experimentally-measured low-lying hadron
spectrum~\cite{Aubin:2004wf,Aoki:2008sm,Durr:2008zz,Bazavov:2009bb,Christ:2010dd,Bernard:2010fr,Gregory:2010gm,Dudek:2011tt,Bietenholz:2011qq,Mohler:2011ke,Gregory:2011sg}, as shown in Fig.~\ref{lqcd:fig:spectrum}.
Lattice-QCD results agree with nonlattice determinations of the charm-and bottom-quark
masses~\cite{Chetyrkin:2009fv,McNeile:2010ji,Beringer:1900zz} and strong coupling~$\alpha_s$
\cite{McNeile:2010ji,Allison:2008xk,Davies:2008sw,Aoki:2009tf,Shintani:2010ph,Bethke:2011tr,Blossier:2012ef}, but now surpass the precision obtained by other methods.
Further, lattice-QCD calculations correctly predicted the mass of the $B_c$
meson~\cite{Allison:2004be,Abulencia:2005usa}, the leptonic decay constants $f_D$ and
$f_{D_s}$~\cite{Aubin:2005ar,Artuso:2005ym}, and the $D\to K \ell \nu$ semileptonic form
factor~\cite{Aubin:2004ej,Widhalm:2006wz} (see Fig.~\ref{lqcd:fig:D2K}) before the availability of precise
experimental measurements.
These successful predictions and postdictions validate the methods of numerical lattice QCD, and demonstrate
that reliable results can be obtained with controlled uncertainties.

In the last five years lattice QCD has matured into a precision tool.
Results with fully controlled errors are available
for nearly twenty matrix elements:
the decay constants
$f_\pi$, $f_K$, $f_D$, $f_{D_s}$, $f_B$ and $f_{B_s}$, 
semileptonic form factors for
$K\to \pi$, $D \to K$, $D\to\pi$, 
$B\to D$, $B\to D^{*}$, $B_s\to D_s$ and $B\to\pi$,
and the four-fermion mixing matrix elements
$B_K$, $f_B^2 B_B$ and $f_{B_s}^2 B_{B_s}$.
By contrast, in 2007, 
only $f_K/f_\pi$ was fully controlled~\cite{whitepaper07}.
The present lattice errors for a sample of matrix elements relevant for the CKM unitarity-triangle fit, along with forecasts for the anticipated lattice errors in five years, can be found in the ``Report of the Quark Flavor Phyics" working group in these proceedings or in Ref.~\cite{USQCD_IF_whitepaper13}.
In the kaon sector, errors are at or below the percent level,
while for $D$ and $B$ mesons errors range from few to $\sim$10\%.  Because these matrix elements cannot be obtained directly from experiment,  it is important to cross-check these results with independent calculations using different lattice actions and analysis methods.  Indeed, this has been done for almost all the quantities noted above. 
This situation has spawned two lattice averaging efforts, 
{\tt latticeaverages.org}~\cite{Laiho:2009eu} and 
FLAG-1~\cite{Colangelo:2010et}, 
which have recently joined forces and expanded to form a worldwide
Flavor Lattice Averaging Group (FLAG-2), 
with first publication expected before the end of 2013.

\subsection{Lattice QCD for the intensity frontier}
\label{subsec:lqcd:IF}

Experiments at the ``intensity frontier" cover a broad range of areas within high-energy, and even nuclear, physics.  The common thread is that, through the use of intense beams and sensitive detectors, they search for processes that are extremely rare in the Standard Model and look for tiny deviations from Standard-Model
expectations.  Therefore the future success of the experimental intensity-physics program hinges on reliable Standard-Model predictions on the same
timescale as the experiments and with commensurate uncertainties.  In many cases, the comparison between the measurements and Standard-Model predictions are currently limited by theoretical uncertainties from nonperturbative hadronic amplitudes that can only be computed with controlled uncertainties that are systematically improvable via lattice QCD.  Thus facilities for numerical lattice QCD are an essential theoretical compliment to the experimental program.
  
In this section we discuss several key opportunities for lattice-QCD calculations to aid in the
interpretation of experimental measurements at the intensity frontier.  In some cases, such as for the determination of CKM matrix elements that are parametric inputs to
Standard-Model predictions, improving the precision of existing calculations is sufficient, and the expected
increase in computing power due to Moore's law will enable a continued reduction in errors.
In other cases, like the muon $g-2$ and the nucleonic probes of non-Standard-Model physics, new hadronic matrix elements
are required; these calculations are typically computationally more demanding, and methods are
under active development.  More details can be found in the USQCD whitepaper ``Lattice QCD at the Intensity Frontier"~\cite{USQCD_IF_whitepaper13}, the document ``Project~X: Physics Opportunities"~\cite{Kronfeld:2013uoa}, and in the summary reports by other working groups in these proceedings. 

\begin{itemize}

\item {\it Quark-flavor physics.} Reducing errors in the hadronic matrix elements involving quark-flavor-changing transitions has been a major focus
of the worldwide lattice-QCD community over the last decade.  The results for some quantities are now very precise, and play an important role in the determination of
the elements of the CKM matrix and in tests of the Standard Model via the global CKM unitarity-triangle fit.

In the kaon sector, errors on gold-plated matrix elements (such as leptonic decay constants and neutral kaon mixing) have been computed to a few percent or better precision, and promising methods are being developed to attack more complicated quantities such as $K\to\pi\pi$ amplitudes~\cite{Blum:2011pu,Blum:2011ng,Blum:2012uk} and the long-distance contributions to the $K_L$-$K_S$ mass difference $\Delta M_K$~\cite{Yu:2011np,Christ:2012se}.  Initial results suggest that calculations of the two complex decay amplitudes $A_0$ and $A_2$ describing the decays $K\to(\pi\pi)_I$ for $I=0$ and 2 respectively are now realistic targets for large-scale lattice QCD
calculations.  The complex $I=2$, $K\to\pi\pi$ decay amplitude $A_2$ has now been computed in lattice QCD with 15\%
errors~\cite{Blum:2011ng,Blum:2012uk}, and a full calculation of $\epsilon^\prime$ with a total error at the 20\% level may be possible in 
two~years.  This advance will open the exciting possibility to search for physics beyond the Standard Model via existing experimental measurements from KTeV and NA48 of direct
$CP$-violation in the kaon system~\cite{Batley:2002gn,Abouzaid:2010ny}.  Further, with this precision, combining the pattern of experimental results for $K\to\pi\nu\bar\nu$ with
$\epsilon'/\epsilon$ can help to distinguish between new-physics models~\cite{Buras:1999da,Kronfeld:2013uoa}.  

The rare kaon decays $K^+ \to \pi^+ \nu \bar{\nu}$ and $K_L \to \pi^0 \nu \bar{\nu}$ are especially promising channels for new-physics discovery because the Standard-Model branching fractions are known to a precision unmatched by any other quark flavor-changing-neutral-current process.  The limiting source of uncertainty in the Standard-Model predictions for $\BR(K^+ \to \pi^+ \nu \bar{\nu})$ and $\BR(K_L \to \pi^0 \nu \bar{\nu})$ is the parametric error from $|V_{cb}|^4,$ and is approximately $\sim$10\%~\cite{Brod:2010hi}.  Therefore a reduction in the uncertainty on $|V_{cb}|$ is essential for interpreting the results of the forthcoming measurements by NA62, KOTO, ORKA, and subsequent experiments at Project~X as tests of the Standard Model.  The CKM matrix element $|V_{cb}|$ can be obtained from exclusive $B \to D^{(*)} \ell\nu$ decays given
lattice-QCD calculations of the hadronic form factors~\cite{Bailey:2010gb}.  In the next five years, the projected improvement in the $B\to D^* \ell\nu$ form factor will reduce the error in $|V_{cb}|$ to
$\lesssim 1.5\%$, and thereby reduce the error on the Standard-Model $K\to\pi\nu\bar{\nu}$ branching
fractions to $\lesssim 6$\%.  With this precision, the theoretical uncertainties in the Standard-Model predictions will be commensurate
with the projected experimental errors in time for the first stage of Project~X.

Errors on $D$- and $B$-meson matrix elements are currently larger than their counterparts in the kaon sector, typically a few to several per cent.  Because $m_c$ and $m_b$ are large relative to typical lattice spacings in current simulations, lattice-QCD simulations of charm and bottom quarks all rely on the use of effective field theory to control the associated discretization errors.  The most well-studied and precise heavy-quark observables are the leptonic decay constants $f_{D_{(s)}}$ and $f_{B_{(s)}}$, for which there are already several independent calculations available using different lattice actions and simulation parameters~\cite{Namekawa:2011wt,Dimopoulos:2011gx,Bazavov:2011aa,Na:2012kp,Na:2012iu,Bernardoni:2012ti,Bazavov:2012dg,Carrasco:2012de,Dowdall:2013tga}.  Lattice calculations of $f_B$ and $f_{B_s}$ are needed for the theoretical predictions of ${\rm BR}(B \to \tau \nu)$ and ${\rm BR}(B_s \to \mu^+ \mu^-)$, respectively, and are currently the largest sources of uncertainty in the Standard-Model rates~\cite{Buras:2012ru}.   With the projected increase in computing resources over the next five years, simulations with finer lattice spacings and physical light-quark masses will enable a straightforward reduction in the errors on $f_B$ and $f_{B_s}$ to the percent level.  Combined with the anticipated improved experimental precision from Belle~II and the LHCb upgrade, this will increase the reach of new-physics searches in these channels.   In the next few years, the errors on other simple hadronic matrix elements will also continue to shrink: examples include the $D \to \pi(K) \ell \nu$ form factors, the $B \to \pi \ell \nu$ form factor, the $B\to D^{(*)}\ell\nu$ form factors, and neutral ${B^0}_{d(s)}$-meson mixing matrix elements.  These will enable even more precise determinations of CKM matrix elements and stringent constraints on the apex of the CKM unitarity triangle.

The scope of lattice-QCD calculations in $B$-meson sector will continue to expand in the next few years.  For example, the branching ratio for $B\to K\ell^+\ell^-$ is now well measured, and increasingly precise results from LHC$b$ and Belle~II are expected.  Present theoretical estimates for the Standard-Model prediction obtain the vector and tensor $B\to K$ form factors from light-cone sum rules.  The first result for these form factors from first-principles lattice QCD appeared recently~\cite{Bouchard:2013eph}, however, and other independent lattice calculations are nearing completion~\cite{Zhou:2012dm}.  The lattice calculation of $B\to K\ell^+\ell^-$ is similar to that of $B\to\pi\ell\nu$ form factor, and similar accuracies are expected over the next five years.  The $B\to K$ vector and tensor form factors are also needed to describe decays involving missing energy, $B\to K X$, in beyond-the-Standard-Model theories~\cite{Kamenik:2011vy}, and analogous form factors are needed for $B\to\pi X$ and  $K\to\pi X$ decays.  Further, the $K\to\pi$ tensor form-factor is needed to evaluate new-physics contributions to $K\to\pi\ell^+\ell^-$~\cite{Baum:2011rm}.  The lattice computations of tensor form factors for $K\to\pi$ and $B\to\pi$ to are straightforward extensions of existing calculations; therefore comparable accuracy to the present errors in the corresponding vector form factors can be achieved soon, and future errors are expected to continue to follow the projections for similar matrix elements.

The $D$-meson system provides complimentary information to the kaon and $B$-meson sectors because charm flavor-changing neutral currents involve down-type quarks.  Experimental measurements of $D^0$-$\overline{D^0}$ mixing and $CP$-violation in $D\to\pi\pi$ and $D\to KK$ decays from LHCb and eventually Belle~II, in particular, offer potentially sensitive tests of the Standard Model and unique probes of new physics, but improved Standard-Model predictions are needed.  Lattice-QCD calculations of the relevant hadronic matrix elements are challenging, and theoretical work towards developing practical methods is underway.  
The calculation of $CP$-violation in $D\to\pi\pi$ and $D\to KK$ decays is more challenging than that for $K\to\pi\pi$ decay, which already represents the present frontier of lattice calculations.  In both cases, one must deal with the fact that two-hadron
states in finite-volume are not asymptotic states.
Further, for $D$ decays, there are many allowed multi-hadron final states with $E< m_D$: $\pi\pi$ and
$K\overline{K}$ mix with $\eta\eta$, $4\pi$, $6\pi$, etc.
The finite-volume states used by lattice QCD are 
inevitably mixtures of all these possibilities, and one
must disentangle
these states so as to obtain the desired matrix element.
Recently, a first step towards 
developing a complete method has been taken~\cite{Hansen:2012tf}, in which the
problem has been solved in principle for any number of two-particle
channels, and assuming that the scattering is dominantly $S$~wave.
This is encouraging, and it may be that this method will allow
one to  obtain semi-quantitative results for the amplitudes
of interest. Turning this method into
practice will likely take $\sim 5$ years due to a number of numerical
challenges (in particular the need to calculate several
energy levels with good accuracy).  In the more distant future, it should possible to generalize the
methodology to include four particle states; several groups
are actively working on the theoretical issues and much progress
has been made already for three particles~\cite{Polejaeva:2012ut,Briceno:2012rv,Guo:2013qla}.
The short-distance contributions to $D^0$-$\overline{D^0}$ to can be calculated via lattice QCD as for kaons and $B$-mesons, but computing the long-distance contributions is more difficult than in $\Delta M_K$ because many more states propagate between the two insertions of the weak Hamiltonian.

\item {\it Neutrino experiments.}  One of the largest sources of uncertainty in accelerator-based
neutrino experiments arises from the determination of the neutrino flux.
This is because the beam energies are in the few-GeV range, for which the interaction with hadronic targets
is most complicated by the nuclear environment.
At the LBNE experiment, in particular, the oscillation signal occurs at energies where quasielastic
scattering dominates.
Therefore a measurement or theoretical calculation of the $\nu_\mu$ quasielastic scattering cross section as
a function of energy $E_\nu$ provides, to first approximation, a determination of the neutrino flux.
The cross section for quasielastic $\nu_\mu n \to \mu^-p$ and $\bar{\nu}_\mu p\to \mu^+n$ scattering is 
parameterized by hadronic form factors that can be computed from first principles with lattice QCD.
The two most important form factors are the vector and axial-vector form factors, corresponding to the $V$ 
and $A$ components of $W^\pm$ exchange.
The vector form factor can be measured in elastic $ep$ scattering.
In practice, the axial-vector form factor has most often been modeled by a one-parameter dipole 
form~\cite{LlewellynSmith:1971zm}
\begin{equation}
    F_A(Q^2) = \frac{g_A}{(1+Q^2/M_A^2)^2},
    \label{lqcd:eq:dipole}
\end{equation}
although other parametrizations have been
proposed~\cite{Kelly:2004hm,Bradford:2006yz,Bodek:2007ym,Bhattacharya:2011ah}.
The normalization $g_A=F_A(0)=-1.27$ is taken from neutron $\beta$~decay \cite{Beringer:1900zz}.
The form in Eq.~(\ref{lqcd:eq:dipole}) in the low~$Q^2$ range relevant neutrino experiments does not rest on a
sound foundation.
Fortunately, the lattice-QCD community has a significant, ongoing effort devoted to calculating $F_A(Q^2)$ 
\cite{Khan:2006de,Yamazaki:2009zq,Bratt:2010jn,Alexandrou:2010hf,Alexandrou:2013joa}.
Recently two papers with careful attention to excited-state contamination in the lattice correlation functions and the chiral extrapolation~\cite{Capitani:2012gj} and 
lattice data at physical pion mass~\cite{Horsley:2013ayv} find values of the axial charge in agreement with experiment, $g_A\approx1.25$.
A~caveat here is that Refs.~\cite{Capitani:2012gj,Horsley:2013ayv} simulate with only $N_f=2$ sea quarks.
If these findings are confirmed by other groups, the clear next step is to compute the 
shape of the form factors with lattice QCD.
If the calculations of the vector form factor reproduce experimental measurements, then one could proceed to 
use the lattice-QCD calculation of the axial-vector form factor in analyzing neutrino data.

Proton decay is forbidden in the Standard Model but is a natural prediction of grand unification.
Extensive experimental searches have found no evidence for proton decay, but future experiments
will continue to improve the limits.
To obtain constraints on model parameters requires knowledge of hadronic matrix elements
$\langle\pi,K,\eta,\ldots|\mathcal{O}_{\Delta B=1} | p \rangle$ of the baryon-number violating operators
$\mathcal{O}_{\Delta B=1}$ in the effective Hamiltonian.
Estimates of these matrix elements based on the bag model, sum rules, and the quark model vary by as much as a factor of three~\cite{Ioffe:1981kw,Claudson:1981gh,Donoghue:1982jm,Martin:2011nd}, and lead to an ${\mathcal O}(10)$ uncertainty in the model predictions for the proton
lifetime. Therefore, \emph{ab initio} QCD calculations of proton-decay matrix elements with controlled systematic
uncertainties of even $\sim 20\%$ would represent a significant improvement, and be sufficiently precise for
constraining GUT theories. Recently the RBC and UKQCD Collaborations obtained the first direct calculation of proton-decay matrix
elements with $N_f=2+1$ dynamical quarks~\cite{Aoki:2013yxa}.
The result is obtained from a single lattice spacing, and the total statistical plus systematic uncertainties
range from 20--40\%.
Use of gauge-field ensembles with finer lattice spacings and lighter pions, combined with a new technique to
reduce the statistical error~\cite{Blum:2012uh}, however, should enable a straightforward reduction of the
errors to the $\sim 10\%$ level in the next five years.

\item {\it Charged-lepton physics.}  Charged-lepton flavor violation (CFLV) is so highly suppressed in the Standard Model that any observation of
CLFV would be unambiguous evidence of new physics.
Many new-physics models allow for CLFV and predict rates close to current limits.
Model predictions for the $\mu \to e$ conversion rate off a target nucleus depend upon the light- and
strange-quark content of the nucleon~\cite{Cirigliano:2009bz}.
These quark scalar-density matrix elements are also needed to
interpret dark-matter detection experiments in which the dark-matter particle scatters off a
nucleus~\cite{Bottino:1999ei,Ellis:2008hf,Hill:2011be}.
Lattice-QCD can provide nonperturbative calculations of the scalar quark content of the nucleon with
controlled uncertainties.
Results for the strange-quark density obtained with different methods and lattice formulations agree at the 1--2$\sigma$ level, and a
recent compilation quotes an error on the average $m_s \langle N| \bar{s}s|N \rangle$ of about
25\%~\cite{Junnarkar:2013ac}.
With this precision, the current lattice results already rule out the much larger values of 
$m_s\langle N|\bar{s}s|N\rangle$ favored by early non-lattice
estimates~\cite{Nelson:1987dg,Kaplan:1988ku,Jaffe:1989mj}.
Lattice-QCD can also provide first-principles calculations of the pion-nucleon sigma
term~\cite{Young:2009zb,Durr:2011mp,Horsley:2011wr,Dinter:2012tt,Shanahan:2012wh} and the charm-quark content
of the nucleon~\cite{Freeman:2012ry,Gong:2013vja}.
A realistic goal for the next five years is to pin down the values of all of the quark scalar density matrix
elements for $q=u,d,s,c$ with $\sim$ 10--20\% uncertainties; even greater precision can be expected on the
timescale of a continuation of Mu2e at Stage~2 of Project~X.

The muon anomalous magnetic moment $a_\mu$ provides one of the most precise tests of the SM and places important constraints on its extensions~\cite{Hewett:2012ns}.
With new experiments planned at Fermilab (E989) and J-PARC (E34) that aim to improve on the current 0.54 ppm measurement at BNL~\cite{Bennett:2006fi} by at least a factor of four, it will continue to play a central
role in particle physics for the foreseeable future.
In order to leverage the improved precision on $g-2$ from the new experiments, the theoretical uncertainty on the Standard Model prediction must be shored-up, as well as be brought to a comparable level of
precision~\cite{Hewett:2012ns}.  

The largest sources of uncertainty in the SM calculation are from the non-perturbative hadronic contributions.
The hadronic vacuum polarization (HVP) contribution to the muon anomaly, $a_{\mu}(\rm HVP)$, has been obtained to a precision of 0.6\% using experimental measurements of $e^{+}e^{-}\to\rm hadrons$ and $\tau\to\rm hadrons$~\cite{Davier:2010nc,Hagiwara:2011af}.
The result including $\tau$ data is about two standard deviations larger than the pure $e^+e^-$
determination, and reduces the discrepancy with the Standard Model to below three standard
deviations~\cite{Davier:2010nc}.
A direct lattice-QCD calculation of the hadronic vacuum polarization with $\sim 1\%$ precision may help shed
light on the apparent discrepancy between $e^{+}e^{-}$ and $\tau$ data;
ultimately a lattice-QCD calculation of $a_{\mu}(\rm HVP)$ with sub-percent precision can circumvent these
concerns.  The HVP contribution to the muon anomalous magnetic moment has been computed in lattice QCD by several groups~\cite{Blum:2002ii,Gockeler:2003cw,Aubin:2006xv,Feng:2011zk,Boyle:2011hu,DellaMorte:2011aa,Burger:2013jya}, and statistical errors on lattice calculations of $a_{\mu}(\rm HVP)$ are currently
at about the 3--5\% level, but important systematic errors remain. Anticipated increases in computing resources will enable simulations directly at the physical quark masses with large volumes, and brute-force calculations of quark-disconnected diagrams, thereby eliminating important systematic errors.

Unlike the case for the HVP, the hadronic light-by-light (HLbL) contribution to the muon anomaly cannot be extracted from experiment. Present estimates of this contribution rely on models~\cite{Prades:2009tw,Nyffeler:2009tw}, and report errors estimated to be 25--40\% range.
Therefore an \emph{ab initio} calculation $a_\mu({\rm HLbL})$ is the highest theoretical priority for $(g-2)_\mu$.
A promising strategy to calculate $a_\mu({\rm HLbL})$ is via lattice QCD plus lattice QED where
the muon and photons are treated nonperturbatively along with the quarks and
gluons~\cite{Hayakawa:2005eq}.
First results using this approach for the single quark-loop part of the HLbL contribution
have been reported recently~\cite{Blum:2013qu}.
Much effort is still needed to reduce statistical errors which remain mostly uncontrolled.
In order to bring the error on the HLbL contribution to, at, or below, the projected experimental uncertainty
on the time scale of the Muon $g-2$ experiment, one must reduce the error on $a_\mu({\rm HLbL})$ to
approximately 15\% or better.
Assuming this accuracy, a reduction of the HVP error by a factor of 2, and the expected reduction in
experimental errors, then the present central value would lie 7--8$\sigma$ from the SM prediction.

\item {\it Tests of fundamental symmetries with nucleons.}  Beyond-the-Standard-Model sources of $CP$ violation that may help to explain the observed baryon asymmetry include nonzero electric dipole moments (EDMs) of leptons and nucleons \cite{Pospelov:2005pr} or neutron-anti-neutron mixing~\cite{Mohapatra:1980qe}.

Lattice QCD can provide first-principles QCD calculations of the strong-$CP$ contribution to the neutron EDM
$d_N/\bar{\theta}$ with improved precision and controlled uncertainties, as well of matrix elements of non-SM
EDM-inducing operators.
Pilot lattice-QCD calculations have already been carried out for this strong-$CP$ contribution to the neutron
and proton EDMs~\cite{Shintani:2008nt,Shintani:2005xg,Aoki:2008gv}.
Currently the statistical errors are still $\sim$30\%, both because of the general property that nucleon
correlation functions have large statistical errors and because the calculation involves correlation functions weighted with the topological charge, which introduces substantial statistical fluctuations.
A lattice-QCD calculation of the matrix elements of dimension-6 operators needed for beyond-the-Standard Model theories is also
underway~\cite{Bhattacharya:2012bf}. This research is still in an early phase, and a reasonable and useful goal for the coming five years is a suite of matrix elements with solid errors at the 10--20\% level.

A low-energy process that would provide direct evidence for baryon number violation from beyond-the-Standard Model physics is the
transition of neutrons to antineutrons, which violates baryon number by two units~\cite{Mohapatra:1980qe}.
A proposed neutron-antineutron oscillation experiment at Project~X could improve the limit on the
$n$-$\bar{n}$ transition rate by a factor of $\sim 1000$.
For many grand unified theories (GUTs) with Majorana neutrinos and early universe sphaleron processes, the
prediction for the oscillation period is between $10^9$ and $10^{11}$ seconds~\cite{Nussinov:2001rb,%
Babu:2008rq,Mohapatra:2009wp,Winslow:2010wf,Babu:2012vc}.
However, this estimate is based on naive dimensional analysis, and could prove to be quite inaccurate when
the nonperturbative QCD effects are properly accounted for.
Calculations of these matrix elements with reliable errors anywhere below 50\% would provide valuable
guidance for new-physics model predictions.
Lattice-QCD calculations can provide both the matrix elements of the six-fermion operators governing this
process and calculate the QCD running of these operators to the scale of nuclear physics.
Initial work on computing these matrix elements is currently underway~\cite{Buchoff:2012bm}.
The main challenge at this stage is to make sufficient lattice measurements to obtain a statistically
significant signal. A first result is expected in the next 1--2 years, with anticipated errors of $\sim 25\%$; results with errors of $\sim 10\%$ or smaller should be achievable over the next five years.

\end{itemize}

\subsection{Lattice field theory for the energy frontier}
\label{subsec:lqcd:EF}

Experiments at the ``energy frontier" directly probe physics up to the TeV scale.  Therefore they can provide unique information about the mechanism of electroweak symmetry breaking realized in nature, either through direct production of new particles or through observing deviations from Standard-Model rate predictions.   Numerical lattice field theory simulations can aid in the search for new physics at current and future high-energy collider facilities in both situations.   

If new TeV-scale resonances are discovered at the LHC or elsewhere, in particular with the same quantum numbers as existing electroweak particles ({\it i.e.} $W'$, $Z'$, and $h'$), these states may be composite objects that result from an underlying strongly-coupled theory such as in Technicolor or Little-Higgs models.   In this case, nonperturbative lattice gauge theory simulations will be needed to make quantitative predictions for the masses and decay constants of these new particles to be compared to the experimental data, and thereby narrow the space of possible new-physics models.  If, on the other hand, non-Standard Model particles are too heavy for direct detection, indirect evidence for Higgs compositeness may still appear as altered rates for electroweak gauge-boson scattering, changes to the Higgs coupling constants, or the presence of additional light Higgs-like resonances.  In this scenario, quantitative lattice-field-theory input may be even more valuable to distinguish between underlying strongly-coupled theories above the TeV-scale that lead to similar experimental observations at lower energies.  

As at the intensity frontier, searches for new physics at high-energy colliders via observing deviations from Standard-Model rates demand precise predictions with controlled uncertainties.  Parametric errors from the quark masses $m_c$ and $m_b$ and the strong coupling constant $\alpha_s$ are the largest sources of uncertainty in the Standard-Model branching-ratio predictions for several Higgs decay channels~\cite{Denner:2011mq}.  Future proposed collider facilities such as the ILC, TLEP, or a muon collider would reduce the experimental uncertainties in Higgs partial widths to the sub-percent level, so reducing the theoretical uncertainties in the corresponding Standard-Model predictions to the same level is essential.  Numerical lattice-QCD simulations provide the only first-principles method for calculating the parameters (quark masses and coupling constant) of the QCD Lagrangian.  Thus supporting lattice-QCD calculations are critical for exploiting precision measurements current and future high-energy colliders.

In this section we discuss key opportunities for lattice gauge theory calculations to aid in the
interpretation of experimental measurements at the energy frontier.  In some cases, such as for the determination of quark masses and $\alpha_s$, precise calculations are already available, and the application of future computing resources to existing lattice methods will enable a continued reduction in errors and further independent cross-checks.  In other cases, like calculations of strongly-coupled beyond-the Standard Model gauge theories, new lattice simulation software and analysis methods are required; these calculations are typically computationally more demanding, and methods are
under active development.  More details can be found in the USQCD whitepaper ``Lattice Gauge Theories at the Energy Frontier"~\cite{USQCD_EF_whitepaper13} and in the summary reports by other working groups in these proceedings. 

\begin{itemize}

\item {\it Parametric inputs $\alpha_s$, $m_c$, and $m_b$ to Standard-Model Higgs predictions.}  The single largest source of error in the theoretical calculation of the dominant Standard-Model Higgs decay mode $H\rightarrow b\overline{b}$ is  the parametric uncertainty in the $b$-quark mass~\cite{Denner:2011mq}.  Because this mode dominates the total Higgs width, this uncertainty is also significant for most of the other Higgs branching fractions.  Parametric uncertainties in $\alpha_s$ and $m_c$ are the largest sources of uncertainty in the partial widths $H\rightarrow gg$ and $H\rightarrow c\overline{c}$, respectively.
 
The most precise known method for obtaining the quark masses $m_c$ and $m_b$ from lattice simulations employs correlation functions of the quark's electromagnetic current~\cite{Allison:2008xk,McNeile:2010ji}  Moments of these correlation functions can easily be calculated nonperturbatively in lattice simulations and then compared to the perturbative expressions which are known to ${\mathcal O}(\alpha_s^3)$.  These moments can also be determined from experimental $e^+e^-$-annihilation data as in Ref.~\cite{Chetyrkin:2009fv}.  The lattice determination of $m_c^{\overline{\rm MS}}(m_c,n_f=4)= 1.273(6)$~GeV is currently the most precise in the world~\cite{Beringer:1900zz}; this is primarily because the data for the lattice correlation functions is much cleaner than the $e^+e^-$ annihilation data.  The uncertainty is dominated by the estimate of neglected terms of ${\mathcal O}(\alpha_s^4)$ in the continuum perturbation theory.  Therefore only modest improvements can be expected without a higher-order perturbative calculation.  

The result for the $b$-quark mass obtained in this way is
$m_b^{\overline{\rm MS}}(m_b, n_f=5) = 4.164(23)$~GeV~\cite{McNeile:2010ji}, and is not currently as precise as the results from
 $e^+e^-$ annihilation~\cite{Chetyrkin:2009fv,Beringer:1900zz}.
The sources of systematic uncertainty  are completely different than for $m_c$.
In this case, perturbative uncertainties are tiny because $\alpha_s(m_b)^4  \ll \alpha_s(m_c)^4$, and discretization errors dominate the current uncertainty, followed by statistical errors.  These should be straightforward to reduce by brute force computing
 power, and so are likely to come down by a factor of two in the next few years, 
 perhaps to $\delta m_b \sim 0.011$~GeV or better.  Precisions of that order for $m_b$ have already been claimed from
 $e^+e^-$ data from
 reanalyses of the data and perturbation theory of Ref.~\cite{Chetyrkin:2009fv}, and coming lattice
 calculations with be able to check these using the computing power expected in the next few years.
 
The strong coupling constant, $\alpha_s$, is also an output of these lattice calculations, and a very
precise value of $\alpha_s(M_Z, n_f=5) = 0.1183(7)$ has been obtained in Ref.~\cite{McNeile:2010ji}, 
with an uncertainty dominated by continuum perturbation theory.
Unlike the heavy-quark masses, for which the correlation function methods give the most precise
results at present, there are numerous good ways of obtaining $\alpha_s$ with lattice methods.  
Several other quantities have been used to make good determinations $\alpha_s$ with lattice QCD, including
Wilson loops~\cite{McNeile:2010ji}, the Adler function~\cite{Shintani:2010ph},
the Schr{\"o}dinger functional~\cite{Aoki:2009tf},
and the ghost-gluon vertex~\cite{Blossier:2012ef}.
All of the lattice determinations are consistent, and each is individually more precise than the most
precise determination that does not use lattice QCD.
The most precise current determination of $\alpha_s$ may improve only modestly over the next
few years, since the error is dominated by perturbation theory.

Lattice-QCD calculations have already determined the quark masses $m_c$ and $m_b$ and the strong coupling $\alpha_s$ more precisely than is currently
being assumed in discussions of Higgs decay channels~\cite{Denner:2011mq}.  The current uncertainties in $\alpha_s$, $m_c$, and $m_b$ from lattice QCD are all currently around a half a per cent and the results, especially for $m_b$, will continue to improve.  For all of these quantities, increased corroboration from independent lattice calculations is expected in the next few years, making the determinations very robust.  If the future lattice error on $\alpha_s$ is reduced by $\sim 30\%$ to $\pm 0.0004$, and that on $m_b$ is reduced by a factor of two to $\pm 0.011~{\rm GeV}$, and these uncertainties are used in the Standard-Model Higgs predictions, then the parametric uncertainty on $\Gamma (H \to b \bar{b})$ would be reduced to 0.8\%, and the total uncertainty on $\Gamma (H \to b \bar{b})$ would be reduced to  2.8\%. 


\item {\it Composite-Higgs model building.}  The recent discovery of the Higgs-like particle at $\sim$ 126~GeV is the beginning of the experimental
search for a deeper dynamical explanation of electroweak symmetry breaking beyond the Standard
Model.  In preparation for the start of the LHC, the lattice-field-theory community has developed an important research direction to study strongly-coupled gauge theories that may provide a natural electroweak symmetry breaking mechanism.  The primary focus of this effort is now on the composite Higgs mechanism, and is described in greater detail in the USQCD white paper ``Lattice Gauge Theories at the Energy Frontier"~\cite{USQCD_EF_whitepaper13}. 

New strongly-coupled gauge theories can behave quite differently than na{\"i}ve expectations based on intuition from QCD.  Applying advanced lattice-field-theory technology to these theories enables quantitative study of their properties, and may provide new nonperturbative insight into this fundamental problem.  The organizing principle of the USQCD program in beyond-the-Standard Model physics is to explore the dynamical implications of (i) approximate scale invariance and (ii) chiral symmetries with dynamical symmetry breaking patterns that give rise to the composite Higgs mechanism with protection of the light scalar mass.  A light composite Higgs that arises as a pseudo-dilaton associated with spontaneous breaking of conformal symmetry may occur in technicolor models, and is also rather natural in supersymmetric theories with flat directions.  Finding an experimentally-viable candidate model requires first identifying a near-conformal ``walking" theory, and then computing the spectrum to see if it contains a light scalar that is well-separated from the remaining new strongly-coupled resonances.   The $S$-parameter in this theory must also be consistent with current precision electroweak constraints.  Once a candidate model is discovered, predictions can be made for experimental observables including the spectrum and modifications to $W$-$W$ scattering that can be tested at the 14-TeV run at the LHC or at future high-energy collider facilities.  A naturally-light composite Higgs can also arise as a pseudo-Nambu Goldstone boson in Little Higgs and minimal conformal technicolor models~\cite{ArkaniHamed:2002qy,Galloway:2010bp}.  At the TeV scale, the physics of the higher-scale theory may be parameterized in terms of an effective theory with a set of low-energy constants whose numerical values are determined by the underlying UV completion.   A central challenge to support this scenario for models based on effective phenomenological Lagrangians is to use lattice field theory to demonstrate that quantitatively viable UV-completions theories exist.  Once a model with a psuedo-Nambu Goldstone Higgs has been established, lattice simulations can provide {\it ab initio} calculations of the low-energy constants from the underlying high-scale theory.  These parameters can then be used to make testable predictions for the 14-TeV LHC run.  In the coming years, lattice calculations of new strongly-coupled gauge theories will become a valuable quantitative tool for narrowing the space of beyond-the-Standard Model theories, and will be essential if the mechanism of electroweak symmetry breaking realized in Nature is nonperturbative.

The development of new tools to study gauge theories beyond QCD has been important to studying both composite-Higgs paradigms.  The existing lattice-QCD software has been extended to enable simulations of theories with arbitrary numbers of colors $N_c$ and flavors $N_f$, and with fermions in the adjoint and two-index symmetric (sextet) representations~\cite{DelDebbio:2010hu,Fodor:2012ni,DeGrand:2013uha}.  Existing lattice methods to study the running coupling in QCD have been extended to identify theories with near-conformal behavior~\cite{Appelquist:2007hu,Appelquist:2009ty,Bilgici:2009nm,Hasenfratz:2010fi,Fodor:2012td}.  Other methods being used to look for viable composite-Higgs theories include computing the mass anomalous dimension (which should be of ${\mathcal O}(1)$ to generate sufficiently large fermion masses without large flavor-changing neutral currents)~\cite{Bursa:2010xr,Appelquist:2011dp,Cheng:2013eu,DeGrand:2013uha} and computing the hadron spectrum to identify the pattern of chiral-symmetry breaking and possible Higgs candidates~\cite{Appelquist:2010xv,Fodor:2012et,Fodor:2012ty,Aoki:2013zsa}.  Calculations of several important low-energy properties such as the $S$ parameter~\cite{Appelquist:2010xv} and $W$-$W$ scattering~\cite{Appelquist:2012sm} have been obtained for a few specific theories, particularly $SU(3)$ gauge theories with increasing numbers of fermions in fundamental and higher representations.  The $S$ parameter in particular is one of the stronger constraints on new physics modifying the
electroweak sector. The lattice result for $S$ in the $SU(3)$ theory with $N_f=2$ fundamental fermions is in conflict with electroweak precision measurements, but the observed reduction in $S$ for $N_f = 6$ fermions indicates that the value of $S$ in many-fermion theories can be acceptably small~\cite{Appelquist:2010xv}, in contrast to more na{\"i}ve scaling estimates~\cite{Peskin:1990zt}.  

\item{\it Supersymmetry model building.}  Supersymmetry (SUSY) is perhaps the most studied extension of the
Standard Model of particle physics. In such models the Higgs is naturally light since it is accompanied by a fermionic partner whose mass is protected by chiral symmetries.   Because the observed low-energy world is not supersymmetric, however, any realistic supersymmetry model must provide a mechanism
for spontaneous SUSY breaking.  Because the SUSY-breaking order parameter cannot belong to any of the supermultiplets, SUSY breaking must arise from interactions with a largely ``hidden" sector in which SUSY is spontaneously broken.   Phenomenologically, this leads to
explicit soft breaking terms in the low-energy Lagrangian of the visible sector.  In general there are many such terms with {\it a priori} undetermined couplings (called the soft parameters), thereby resulting in a large parameter space and a lack of predictivity at low energies.  The particle spectrum of the visible sector is often assumed to be independent of the details of the hidden sector, and only affected by the mechanism for mediating the breaking.  When the hidden sector is strongly-coupled, however, predictions for the visible sector depend sensitively on the SUSY-breaking mechanism~\cite{Cohen:2006qc,Murayama:2007ge}.

Dynamical SUSY breaking generated via dimensional transmutation in the hidden sector provides a natural mechanism for the large separation between the Planck scale and the expected supersymmetry breaking scale of a few TeV~\cite{Witten:1981nf}.  A simple candidate for this hidden sector is a supersymmetrized version of QCD with $N_c$ colors and $N_f$ massive flavors~\cite{Intriligator:2006dd}.  Nonperturbative lattice field theory methods can be used to study the vacuum structure and dynamics of super QCD, and ultimately to compute the values of the numerous soft parameters in the low-energy theory in terms of only a handful parameters of the hidden-sector super-QCD theory (given the mechanism for mediating SUSY breaking).  This will place important quantitative constraints on low-energy supersymmetric models (such as the MSSM) and provide essential quantitative input to realistic supersymmetric model building.  Recent lattice efforts have focused on super-Yang Mills with the gauge group $SU(2)$ and ${\mathcal N}=1$ supersymmetries~\cite{Giedt:2008xm,Endres:2009yp,Demmouche:2010sf}, and shown that the obtained value of the gluino condensate agrees with theoretical expectations.  Building a realistic SUSY-breaking sector capable of yielding the soft parameters of the low-scale SUSY theory, however, requires adding quarks (plus their scalar superpartners) and extending the gauge group to a larger number of colors.  The simplest super-QCD system that is expected to exhibit metastable SUSY-breaking vacua corresponds to four colors and five flavors~\cite{Intriligator:2006dd};  the simulation of such a theory with current algorithms will require petascale computing resources, and therefore is not anticipated for the near future.  Nevertheless, further numerical study of super QCD using lattice methods will help to develop the necessary software and analysis tools for future more realistic simulations, and will likely lead to valuable insights regarding dynamical SUSY breaking along the way to the more ambitious goal of computing the soft-breaking parameters from the underlying hidden-sector super-QCD Lagrangian.

\end{itemize}

\section{Resources for lattice studies at the energy and intensity frontiers}
\label{sec:lqcd:resources}

In this section we discuss the computational and software infrastructure resources needed to reach the scientific goals set out above.
We focus on the efforts and plans in the US, but comparable efforts are ongoing in Europe and Japan.  We begin in Sec.~\ref{subsec:lqcd:current} by describing the computational resources currently used by the U.S. lattice gauge theory community.  Then, in Sec.~\ref{subsec:lqcd:fiveyear}, we provide forecasts for computing capabilities on an approximately five-year time scale based on extrapolations from current processor and networking roadmaps.  We also discuss the new physics calculations that will be enabled with these resources.  In Sec.~\ref{subsec:lqcd:future} we speculate on possible computing resources beyond 2018.

\subsection{Computational requirements and current resources}
\label{subsec:lqcd:current}

The simulation codes of lattice gauge theory require substantial computing
resources in order to calculate hadron masses and interactions with sufficient
precision to test the Standard Model against emerging experimental
measurements.  At present, lattice theorists in the United States run these
codes on a variety of hardware, including large supercomputers at DOE and NSF
leadership computing facilities, medium-range supercomputers and clusters at
centers supported by the NSF's XSEDE Program and at NERSC, dedicated computing systems at
Fermilab, Jefferson Lab, and Brookhaven, and smaller clusters at university
sites.  Depending upon the size of problem and the type of calculation, either
{\em capability} or {\em capacity} computing systems are required.  Access to
both types of computing resources is essential for the lattice field theory
community.

Lattice gauge theory simulations require parallel programming techniques, with
the calculations running cooperatively across hundreds to many thousands of
processors or processor cores.  The simulations must be run on hardware suitable
for massively parallel computations.   Although the simulations are
floating point intensive, on all current high-performance computing systems
throughput is limited by the rate that operands can be supplied to the
floating point execution units, either because of memory bandwidth limitations
or by the latency and bandwidth of interprocessor communications.
Interprocessor communications of data rely on message-passing
algorithms, typically implemented using an MPI~\cite{MPI} library.

At present, lattice theorists in the United States run their codes on a
variety of high-performance computing hardware.  The first type is commodity clusters based on Intel or AMD x86
processors and Infiniband networks, which have hundreds of nodes and
thousands of cores.  A second type is accelerated commodity clusters, similar to the the
standard clusters but with general purpose graphics processing units (GPUs) or
Intel many integrated core (MIC) accelerators installed in each server; these
clusters have fewer nodes but typically hundreds of accelerators.  A third type is very
large scale Cray supercomputers, consisting of thousands of AMD x86 processors
with a proprietary network, with the newest models also containing thousands
of GPUs.  Finally, lattice theorists use very large scale IBM BlueGene supercomputers, consisting of
hundreds of thousands of PowerPC cores interconnected on a proprietary
network.

The Cray and IBM supercomputers provide better scaling for large
processor-count jobs and so are capable of efficiently performing calculations
using tens of thousands of processors.  Computing allocations on these DOE
leadership-class facilities favor calculations that require significant
fractions of the machines, and operational policies restrict running jobs to
those with large minimum processor counts.  Commodity clusters and accelerated
clusters give better cost efficiency for smaller jobs, spanning hundreds to a
few thousand core.
More than half of integrated computing capacity used by USQCD
since 2006 has been provided by commodity clusters and accelerated clusters.

Prior to 2006, computing capacity improvements in new generations of
processors resulted from the higher clock speeds enabled by smaller
semiconductor feature sizes.  Device physics factors slowed the rate of
improvement as leakage currents and other inefficiencies began to dominate
power consumption.  Processor manufacturers instead moved to putting multiple
processing cores in each processor package.  As feature sizes decreased, in
the same die area these multicore processors first provided two, then four,
and now as many as sixteen cores.  The cores within a package are coupled
together with shared memory caches and common memory address spaces.  For lattice-QCD
software, which already coped via message passing with the difficulties of
running a single calculation across a large number of cooperating processors,
the transition to multicore processors was largely seamless.  One significant
issue with multicore systems did emerge: the need to strictly assign processes
to individual cores and to use ``local'' memory; all recent commodity systems
with multiple processor sockets use NUMA (Non-Uniform Memory Access) designs
that cause additional latency and effectively lower memory bandwidth when a
core accesses main memory attached to a different processor socket.   

Since 2006, lattice theorists began to exploit general purpose graphics
processing units (GPUs) for their floating point computing capabilities.  Two
major vendors, NVIDIA and ATI, began selling numerical accelerators based on
their GPU chips and provided programming environments such as CUDA (NVIDIA)
and OpenCL (ATI and others).  Unlike the multicore processors discussed above,
which can be utilized with conventional message-passing software, these
accelerators have orders of magnitude higher core counts, with much smaller
local memory per core.  Fully exploiting the potential computing capacity of
these accelerators requires considerable expertise and programming effort.
For large lattice-QCD simulations, the problems must be spread across multiple
accelerators, leading to a hierarchy of computing resources with very
different performance: very fast but small local memory physically attached to
the accelerator cores, much larger but slower (higher latency, lower
bandwidth) memory in the host system housing the accelerator, and additional
processors, accelerators, and memory in host systems connected via an
Infiniband network.

The generation of ensembles of gauge-field configurations is the largest
numerical problem in lattice-QCD simulations.  Because the configurations in an
ensemble are members of a Markov chain, the individual jobs must be performed
in series.  To minimize time-to-solution, the individual calculations must run
on a large number of cores on machines with architectures that deliver
excellent scaling at very high processor count.  Typical jobs span tens of
thousands of cores, running for months to produce the order thousand
configurations of a given ensemble.  Such calculations require capability
machines such as Blue Gene or Cray supercomputers.

Capability hardware operated by the DOE and currently employed for lattice-QCD
calculations include BlueGene/P and /Q, and Cray XK7 supercomputers.  The
USQCD collaboration applies as a single entity for INCITE allocations on the
DOE leadership-class computing facilities (LCF) at Argonne National Lab
(BlueGene hardware) and Oak Ridge National Lab (Cray hardware).  These
facilities are funded by the DOE Advanced Scientific Computing Research (ASCR)
program office.  The USQCD INCITE awards for lattice-QCD simulations are
consistently among the largest allocations at either of the ANL or ORNL LCF.
For calendar year 2012, these allocations were 50M and 46M core hours,
respectively, and in 2013, 290M (250M on the BlueGene/Q and 40M on the
BlueGene/P) and 140M core-hours.  In terms of integrated sustained teraflops,
for lattice-QCD applications 290M and 140M core-hours correspond to
approximately 90 and 22~Tflop/sec-yrs.

Computations of operator expectation values using gauge-field ensembles are
relatively more I/O-intensive than ensemble generation, span a wide range of
smaller job sizes, and can be run in parallel.  That is, such analysis
calculations, repeated for each member of an ensemble, can consist of tens to
hundreds of independent jobs running simultaneously on a large capacity
computing system.  Time-to-solution is not critical for the individual jobs,
so even though they execute many of the same computational kernels as the
ensemble generation calculations on capability machines, they can be run using
fewer processors; because of strong scaling effects, running at lower
processors counts yields better performance per processor on the kernels.
Depending upon the specific stage of the analysis computation, the individual
jobs range in size from requiring a single multicore computer to many
thousands of cores across hundreds of computers.  A single analysis campaign,
such as the calculation of a leptonic decay constant, may consist of tens of
thousands of individual jobs.  These jobs run most efficiently and cost
effectively on large commodity clusters.

Capacity hardware consists of Infiniband-coupled commodity clusters, some of
which include GPU hardware
accelerators.  The DOE HEP and NP program offices have supported the lattice
gauge theory community by funding since FY2006 dedicated capacity hardware at
Fermilab, Jefferson Lab, and Brookhaven.  Lattice gauge simulations for
nuclear physics have similar computational requirements to those for high
energy physics and can utilize the same hardware.  Two joint HEP/NP projects,
LQCD (FY06-FY09) and LQCD-ext (FY10-FY14), have provided funds for hardware
purchases and support personnel.  USQCD has submitted a proposal for a project extension, LQCD-ext~II, which would run from FY14-FY19.  These dedicated capacity
resources are allocated by USQCD.  As of the beginning of July 2013, the
dedicated USQCD hardware at Fermilab, Jefferson Lab, and Brookhaven has a
total capacity of 570M and 770M core-hours, respectively, on conventional and
GPU-accelerated hardware.  In terms of integrated sustained teraflops, these
correspond, respectively, to 88~Tflop/sec-yrs and 119~Tflop/sec-yrs.
Because sufficient allocations are not available via INCITE, and because the LCFs require that individual jobs use a large fraction of the computers, these dedicated
USQCD hardware resources provide essential computing capacity.
In Table~\ref{tab:current} we list the LCF capability and dedicated
capacity resources utilized for lattice-QCD simulations since 2010.
The capability resources are broken out showing both the ANL and ORNL
leadership-class facilities; the capacity resources include all usage  on the 
DOE HEP and NP funded hardware at Fermilab, Jefferson Lab, and BNL.

\begin{table}[t]
\begin{center}
\caption{Utilized core-hours of leadership-class facility
(LCF) and dedicated capacity hardware for lattice-QCD simulations.  The
conversion factors for lattice-QCD sustained Tflop/sec-years, assuming 8000
hours per year, is 1 Tflop/sec-year = 3.0M core-hour on BlueGene/Q hardware, and 1 Tflop/sec-year = 6.53M core-hour on
BlueGene/P and Cray hardware. Only USQCD-Collaboration resources are shown.
The drop in ANL LCF utilized capacity in 2012 occurred
because fewer opportunistic core-hours (``zero-priority queues'') were
available due to increased demand by other facility users.
\vspace{1.5mm}}
\label{tab:current}
\begin{tabular}{lccc}  
\hline\hline
Year & ANL LCF & ORNL LCF & Dedicated Capacity Hardware \\[-0.75mm]
& (BG/P + BG/Q core-hours) & (Cray core-hours) & (core-hours) \\[0.5mm]  \hline
2010 & 187M & 53.6M & 125M \\
2011 & 182M & 49.8M & 205M \\
2012 & 143M & 77.9M & 330M \\
2013 & 290M (allocated) & 140M (allocated) & 971M (planned) \\ \hline\hline
\end{tabular}
\end{center}
\end{table}

Non-DOE supported resources are also used for lattice-QCD calculations.  USQCD
has a PRAC grant for the development of code for the NSF's petascale computing
facility, Blue Waters, and has a significant allocation on this
computer during 2013.  Subgroups within USQCD also make use of computing
facilities at the DOE's National Energy Research Scientific Computing Center
(NERSC), the Lawrence Livermore National Laboratory (LLNL), and centers
supported by the NSF's XSEDE Program.  In addition, the RBC Collaboration has
access to dedicated Blue Gene/Q computers at the RIKEN BNL Research Center at
Brookhaven National Laboratory and the University of Edinburgh.

Because of the variety of processor types and parallel architectures,
efficient utilization of the above computing resources requires flexible and
effective software.  Since 2004, DOE grants to USQCD during the three
SciDAC~\cite{SciDAC} programs (2001-2006, 2006-2011, and 2011-2016) led to the
development of the USQCD software stack~\cite{SciDAC-software}.  This stack
includes low-level communications and I/O application program interfaces
(APIs) implemented via libraries ported to and optimized for each of the
architectures.  The stack includes linear algebra libraries with routines that
operate on single lattice sites, or across a full lattice with communications
between neighboring sites.  Lattice-QCD applications 
utilize the various libraries of the software stack to run efficiently on any
of the available computing resources.  The USQCD software stack is a publicly
available resource supporting all of the main lattice gauge and fermion
actions in current use.  Further, it provides a general purpose framework that
can be extended to other quantum field theories besides QCD.

In addition to the computational resources discussed above, lattice field
theory simulations require data storage and network resources.  Tape archives
of the various gauge-field ensembles are maintained at multiple sites to
insure against accidental loss.  Ensembles reside on disk at any of the
computational resource sites where they are required as input data for
analysis jobs.  Intermediate data products generated and used in the analysis
campaigns, such as quark propagators, reside on disk.  Computationally
expensive propagators may be archived to tape, as well as distributed to any
other site where they serve as inputs to additional analyses.  Although the
volumes are modest compared to the various experiments, lattice field theory
data require substantial storage resources.  For example, at Fermilab as of
July 2013 over 1.8 petabytes of lattice field theory data were on tape, an
increase of 0.8 petabytes over the volume stored as of July 2012.  The
dedicated capacity clusters at Fermilab and Jefferson Lab utilize more than
1.4 petabyes of disk configured in parallel file systems.  Fortunately, USQCD
can leverage the storage infrastructure at Fermilab and Jefferson Lab built to
support the DOE experimental programs.  The transfer of data between sites
relies on national data networking infrastructure, such as ESnet, to achieve
the required data rates.  Similar to data storage, the volume and rate of data
transfer is modest compared to that required by experiments, but is
nevertheless substantial.

By matching the scale of a lattice field theory simulation to the most
suitable type and size of machine, the USQCD community maximizes the overall
cost-effectiveness of available computing resources.  Utilizing both
leadership-class facilities and dedicated clusters is an effective means for
meeting the computing needs of the lattice community.  Continued support of
both the national supercomputing centers and of dedicated USQCD hardware, and
support for software and algorithm development, will be needed to meet the
scientific goals enumerated in Secs.~\ref{subsec:lqcd:IF}
and~\ref{subsec:lqcd:EF}.

\subsection{Projected capabilities in the next five years}
\label{subsec:lqcd:fiveyear}

According to publicly available information from vendor roadmaps, during the
next several years commodity server computers, such as those used in the
dedicated lattice-QCD clusters discussed in Section~\ref{subsec:lqcd:current}
above, will likely be similar to the current multicore machines.  Core counts
per processor will increase two-fold or more, improvements in core
architectures will yield higher flop counts per clock cycle, and memory
bandwidth will increase through higher memory clock speeds and perhaps through
new memory architectures.  Infiniband or similar networking will continue to
increase in speed, doubling every three to five years.  Together, these
hardware improvements should continue the exponential drop in cost per
sustained flop for lattice-QCD simulations that has been observed for more
than the past decade.  However, during the two dedicated hardware projects
that ran from FY2006 to FY2009 and from FY2009-FY2014, the halving time for
price per flop has increased, from about 19 months to over 26 months.

Better cost effectiveness, with the burden of more difficult programming, will
continue to be available by incorporating accelerators such as GPUs and the
emerging Intel MIC hardware into conventional servers.  These accelerators
will continue the trend of having large low power core counts with small, fast
local memory, with challenging communications bottlenecks for accessing
non-local memory and communicating with neighbor processors and accelerators
via the system I/O bus (``PCI express'') and over Infiniband or similar
networks.  
 
Judging from the trends of the last half decade, the evolution of
supercomputers at the leadership-class facilities may follow two distinct
patterns: one similar to that observed on the IBM BlueGene series of machines,
and the other to the Cray series.  On BlueGene-like hardware, future machines
may have increased core counts and hardware threads per node, with each
processor continuing to run at low speeds compared with Intel and AMD x86
processors.  For large core-count jobs, this requires heterogeneous software
which simultaneously uses multithreading across local cores and hardware
threads, and message passing between nodes.    The current BlueGene/Q
supercomputer exemplifies this pattern.  Software developed under the
current SciDAC-3 lattice-QCD grant copes well with this architecture.

On Cray-like hardware, future machines may have a larger fraction of their
computing capacity delivered by accelerators such as NVIDIA GPUs or by
future Intel MIC-architecture coprocessors.  These processing elements would
have very high core counts, perhaps in the thousands, running at low power and
with access to small, fast local memory.  Problems of interest for lattice
field theory would have to be spread across thousands of these coprocessors
There would be challenging communications bottlenecks between the GPU or
other coprocessor elements, with access to non-local memory only with high
latency and low bandwidth over the local I/O bus for neighbor elements, and
over the high-performance network for more distant coprocessors.  The Titan
supercomputer at ORNL exemplifies this pattern.  To fully exploit
supercomputers like Titan, new algorithms must be developed to minimize the
penalties imposed by the communications bottlenecks.  Recently~\cite{QUDAdd}
an implementation of L\"{u}scher's~\cite{DomainDecomp} domain decomposition
communications-avoidance algorithm developed with DOE support via the SciDAC
program has demonstrated very good strong scaling using hundreds of GPUs.

As computational capacities increase over the next five years, data storage
volumes will increase as well.  Gauge-field ensemble and propagator data file
sizes will increase at the planned finer lattice spacings and larger
simulation volumes.  Since these are space-time simulations, data volumes
increase as the fourth power of inverse lattice spacing.  In 2013, the
LQCD-ext dedicated hardware project increased the fraction of the hardware
budget spent on storage from five to eight percent to accomodate the increased
demand.  The fraction of time spent by analysis jobs on file I/O will continue
to increase.  Improvements to software will be necessary to optimize I/O, and
the workflow patterns employed during analysis campaigns may need to change to
reduce demands on disk and tape storage.

The advent of petascale supercomputers is for the first time enabling
widespread simulations with physical up and down quark masses at small lattice
spacings and large volumes.  This development will enable major advances on a
range of important calculations.  Over the next five years, the US lattice-QCD
effort in precision matrix elements for the intensity frontier will generate
large sets of gauge-field ensembles with the domain-wall fermion
(DWF)~\cite{Kaplan:1992bt,Furman:1994ky,Vranas:2006zk} and highly improved
staggered quark (HISQ)~\cite{Follana:2006rc} lattice actions. Each of these
formulations has its own advantages, and the availability of two independent
sets of configurations will enable valuable cross-checks on lattice
calculations of the most important quantities.

While the challenges to further reductions in errors depend on the quantity, a
few key advances in the next five years will help a broad range of
calculations.  First, the widespread simulation of physical $u$ and $d$ quark
masses will obviate the need for chiral extrapolations.  Such simulations have
already been used for studies of the spectrum and several matrix elements
including the leptonic decay constant ratio $f_K/f_\pi$ and the neutral kaon
mixing parameter
$\hat{B}_K$~\cite{Aoki:2009ix,Durr:2010vn,Durr:2010aw,Bazavov:2013cp,Dowdall:2013rya}.
A second advance will be the systematic inclusion of isospin-breaking and
electromagnetic (EM) effects.  Once calculations attain percent-level
accuracy, as is the case at present for quark masses, $f_K/f_\pi$, the
$K\to\pi\ell\nu$ and $B\to D^*\ell\nu$ form factors, and $\hat B_K$, one must
study both of these effects.  A partial and approximate inclusion of such
effects is already made for light-quark masses, $f_\pi$, $f_K$ and $\hat B_K$.
Full inclusion would require nondegenerate $u$ and $d$ quarks and the
incorporation of QED into the simulations, both of which are planned for the
five-year DWF and HISQ configuration-generation programs.  A final
across-the-board improvement that will likely become standard in the next five
years is the use of charmed sea quarks.  These are already included in two of
the major streams of gauge-field ensembles being generated
worldwide~\cite{Baron:2009wt,Bazavov:2012xda}.

The anticipated increase in computing resources over the next five years will
significantly benefit the already mature quark-flavor physics program,
improving the precision of weak-matrix elements needed to determine CKM matrix
elements, constrain the CKM unitarity triangle, and search for evidence of
non-Standard Model quark flavor-changing interactions.  It will also enable
dramatic reduction in the errors of nucleon matrix elements needed to compute
nucleon-neutrino scattering cross sections, interpret $\mu \to e$ conversion
and dark-matter experiments, and search for violations of fundamental
symmetries of the Standard Model.  Lattice calculations involving nucleons,
however, typically require larger spatial volumes and more statistics than
their meson counterparts.  Therefore achieving comparable percent-level
precision for nucleon matrix elements will require more computing time than
the USQCD anticipates receiving on the leadership-class machines and on
dedicated hardware in the next few years, so the US lattice-QCD community
could profitably take advantage of additional computing resources were they to
become available.

\begin{table}[t]
\begin{center}
\caption{Available resources for lattice-QCD simulations assumed for the planned program of physics calculations.  The conversion factors for lattice-QCD sustained
Tflop/sec-years, assuming 8000 hours per year, is 1 Tflop/sec-year = 3.0M core-hour on BlueGene/Q hardware, and 1 Tflop/sec-year = 6.53M core-hour on BlueGene/P and Cray hardware. \vspace{1.5mm}}
\label{tab:fiveyear}
\begin{tabular}{lccc}
\hline\hline  
Year & Leadership Class  & Dedicated Capacity Hardware  \\[-0.75mm] 
& (Tflop/sec-yrs) & (Tflop/sec-yrs) \\[0.5mm] \hline
2015 & 430 & 325 \\
2016 & 680 & 520 \\
2017 & 1080 & 800 \\
2018 & 1715 & 1275 \\ 
2019 & 2720 & 1900 \\ \hline\hline
\end{tabular}
\end{center}
\end{table}

The planned U.S. physics program over the next five years is described in detail
in the USQCD
whitepapers~\cite{USQCD_EF_whitepaper13,USQCD_IF_whitepaper13,USQCD_NP_whitepaper13,USQCD_Thermo_whitepaper13}.
This physics program assumes the availability to USQCD of capability resources
at the DOE leadership-class facilities, as well as the availability of dedicated
capacity resources at Fermilab, Jefferson Lab, and BNL, deployed and operated
under the proposed LQCD-ext~II project extension.  The sustained LQCD Tflop/sec-years provided by
these resources by year are given in Table ~\ref{tab:fiveyear}. In all, this
program of physics calculations 
will require well over an order of magnitude of increased computing capacity
beyond that used in prior years.  Further, over an order of magnitude increase
in storage utilization (disk and tape) from the current approximately 2
petabyte usage will be needed to support the simulations.   This computing and
storage capacity can be provided by the growth
of the various leadership-class facilities and larger allocations on those
supercomputers, and by the continued availability and expansion of the
dedicated hardware for lattice field theory supported by the DOE HEP and NP
program offices.  The anticipated evolution of high-performance computing
hardware will also require the evolution of software and the introduction and
refinement of new techniques and algorithms.  DOE support for the personnel to
invent and refine algorithms and to provide new software will also be necessary to
exploit the hardware and to complete the planned physics program.

\subsection{Lattice computing beyond 2018}
\label{subsec:lqcd:future}

Beyond the five-year timescale, concrete projections for the physics
capabilities and computing needs of numerical lattice calculations become less
reliable and more speculative.  Lattice field theory is a theoretical area of
research, and the development of new lattice formulations and analysis methods
as well as better computing algorithms drive rapid, but
difficult-to-anticipate, evolution of the field.  Here we attempt to
extrapolate to the extended time period covered by the Snowmass study based on
existing lattice methods and increased computing resources.  For concreteness,
we focus on weak matrix-element calculations, for which current lattice
results are the most precise and there is the most quantitative experience.

We begin with the conservative assumptions that exascale performance
($10^{18}$ flops/second) will be achieved by 2022, and that a further factor
of 100 will be available by 2032.  Present large-scale lattice calculations at
physical quark masses are performed in volumes of linear size $L \approx 6$ fm
and with inverse lattice spacing $1/a$ as small as $\sim 2.5$ GeV.  Thus,
these $10^2$ and $10^4$ advances in computer capability will allow an increase
in physical volume to 15 and 36 fm or in inverse lattice spacing to 5 and 10
GeV, respectively.  Statistical errors can be reduced by a factor of ten, or
even one hundred, as needed.  These three directions of substantial increase
in capability translate directly into physics opportunities.  The large
increase in possible Monte Carlo statistics will enable a reduction in the
errors on many nucleon matrix elements to the percent level, and on quark
flavor-changing matrix elements to the sub-percent level.  Such increased
statistics will also directly support perhaps few-percent precision for
results that depend on quark-disconnected diagrams such as $\epsilon'$ and the
$K_L-K_S$ mass difference.  For most QCD calculations, the non-zero pion mass
implies that finite volume effects decrease exponentially in the linear size
of the system.  However, this situation changes dramatically when
electromagnetic effects are included.  Here the massless photon leads to
substantial finite volume errors which decrease only as a power of $L$ as the
linear system size $L$ becomes large.  The ability to work on systems of
linear size 20 or 30 fm will play an important role in both better
understanding electromagnetic effects using lattice methods, and achieving the
10\% errors in the computation of such effects that are needed to attain 0.1\%
overall errors in quantities such as the light-quark-mass ratio $m_u/m_d$ and
the leptonic decay-constant ratio $f_K/f_\pi$.  Finally the ability to work
with an inverse lattice spacing as large as 10 GeV will allow substantial
improvements in the treatment of charm and bottom quarks, and enable
determinations of many quantities involving $B$ and $D$ mesons with errors
well below 1\%.

Clearly an enhanced computational capability of four orders of magnitude,
coupled with possibly equally large theoretical and algorithmic advances, will
have a dramatic effect on the phenomena that can be analyzed and precision
that can be achieved using lattice methods.  The possibility of making
Standard-Model predictions with errors that are an order-of-magnitude smaller
than present experimental errors will create an exciting challenge to identify
quantities where substantially increased experimental precision is possible.
With the ability to make highly accurate Standard-Model predictions for a
growing range of quantities, experiments can be designed to target those
quantities that are potentially most sensitive to physics
beyond-the-Standard-Model, rather than being limited to those quantities which
are least obscured by the effects of QCD.

\section{Summary}
\label{sec:lqcd:summ}

Facilities for numerical lattice gauge theory are an essential theoretical compliment to the experimental
high-energy physics program.  Lattice-QCD calculations now play an essential role in the search for new physics at the intensity frontier.
They provide accurate results for many of the hadronic matrix elements needed to realize the potential of
present experiments probing the physics of flavor. The methodology has been validated by comparison 
with a broad array of measured quantities, several of which had not been well measured in experiment when the first good lattice calculation became available.  In the next decades, lattice-QCD has the welcome opportunity to play an expanded role in the search for new physics at both the energy and intensity frontiers.  

The USQCD Collaboration, which consists of most theoretical physicists in the U.S. involved in the numerical study of QCD and beyond-the-Standard Model theories using lattice methods, has laid out an ambitious vision for future lattice calculations matched to the experimental priorities of the planned experimental high-energy physics program in the white papers ``Lattice QCD at the Intensity Frontier" and ``Lattice Gauge Theories at the Energy Frontier" \cite{USQCD_IF_whitepaper13,USQCD_EF_whitepaper13}.  These detailed documents present a concrete five-year plan for both the collaboration's foremost scientific goals and the theoretical, algorithmic, and computational strategies for achieving them.

In the U.S., the effort of the lattice gauge-theory community has been supported in an essential way by hardware and software support provided to the USQCD Collaboration.  The USQCD Collaboration's hardware project is up for renewal in 2015, and USQCD is currently in the midst of obtaining CD-0 approval from the DOE for the project extension LQCD-ext~II.  Achieving the goals outlined in these white papers and meeting the needs of current, upcoming, and future experiments will require continued support of both the national supercomputing centers and of dedicated USQCD hardware through LQCD-ext~II, investment in software development through SciDAC funding, and support of postdoctoral researchers and junior faculty through DOE and NSF grants to lab and university lattice gauge theorists.   

The main findings of this report are summarized here:

\begin{itemize}

\item The scientific impact of many future experimental measurements at the energy and intensity frontiers hinge on reliable Standard-Model predictions on the same time
scale as the experiments and with commensurate uncertainties. Many of these predictions require nonperturbative hadronic matrix elements or fundamental QCD parameters that can only be computed numerically with lattice-QCD. The U.S. lattice-QCD community is well-versed in the plans and needs of the experimental high-energy program
over the next decade, and will continue to pursue the necessary supporting theoretical calculations.   Some of the highest priorities are improving calculations of hadronic matrix elements involving quark-flavor-changing transitions which are needed to interpret rare kaon decay experiments, improving calculations of the quark masses $m_c$ and $m_b$ and the strong coupling $\alpha_s$ which contribute significant parametric uncertainties to Higgs branching fractions, calculating the nucleon axial form factor which is needed to improve determinations of neutrino-nucleon cross sections relevant experiments such as LBNE, calculating the light- and strange-quark contents of nucleon which are needed to make model predictions for the $\mu \to e$ conversion rate at the Mu2e experiment (as well as to interpret dark-matter detection experiments in which
the dark-matter particle scatters off a nucleus), and calculating the hadronic light-by-light contribution to muon $g-2$ which is needed to solidify and improve the Standard-Model prediction and interpret the upcoming measurement as a search for new physics.  Lattice field-theory calculations will also increasingly contribute to collider experiments at the LHC 14-TeV run by providing quantitative nonperturbative input for Higgs and other new-physics model building.

\item The successful accomplishment of USQCD's scientific goals requires access to both capacity and capability machines, and hence support for both leadership-class facilities and dedicated computing clusters.  Use of leadership-class facilities alone would provide insufficient computational resources needed to complete the planned calculations, and would be unsuitable for the mix of lattice-field-theory job requirements.   USQCD's experience and proven track-record with purchasing, deploying, utilizing, and maintaining dedicated clusters will enable the collaboration to take advantage of future improvements in commodity clusters, such as increased core counts per processor and improved memory and networking bandwidth.   USQCD's five-year computing strategy uses current vendor roadmaps to anticipate the probable evolution of high-performance computing hardware over this time period.  The purchase of new dedicated lattice hardware on an annual basis, however, provides essential flexibility to accommodate changes and developments, and thereby to purchase the most cost-effective machines for lattice-field-theory calculations.

\item The successful utilization of future computing resources requires software that runs efficiently on new computing architectures, and hence support for postdocs and scientific staff to develop lattice-gauge-theory code.  Such positions cannot be supported by grants to lab and university theory groups alone.   The USQCD Collaboration's libraries for lattice-field-theory calculations are publicly available and are used by most of the U.S. community.  USQCD's experience and proven track-record in developing software for diverse machines such as IBM and Cray supercomputers, PC commodity clusters, and GPU-accelerated clusters, will enable the collaboration to fully exploit the computing capacity of future architectures.  

\item Support of USQCD through hardware and software grants, access to leadership-class computing facilities, and funding lab and university theorists, is essential to fully capitalize on the enormous investments in the DOE's high-energy physics and nuclear-physics experimental programs.  Given continued support of the lattice-gauge-theory effort in the U.S. and worldwide, lattice calculations can play a key role in definitively establishing the presence of physics beyond the Standard Model and in determining its underlying structure.

\end{itemize}


\end{document}